\documentclass[fleqn,10pt]{wlscirep}

\usepackage{bm} 
\usepackage{amsmath}
\usepackage[english]{babel}
\usepackage[utf8]{inputenc}
\usepackage{dsfont}
\usepackage{graphicx}
\usepackage{lipsum}
\usepackage{bibentry}
\usepackage{xcolor}         

\newcommand{\Title}{Uhlmann number in translational invariant systems}
\newcommand{\HH}{\mathcal{H}}
\newcommand{\daga}[1]{#1^{\dagger}}

\newcommand{\Tr}{\text{Tr}}
\newcommand{\munu}[1]{#1_{\mu \nu}}
\newcommand{\Ch}{\text{Ch}}
\newcommand{\funk}[1]{#1 ( \mathbf{k} ) }
\newcommand{\pedk}[1]{#1_{ \mathbf{k} } }
\newcommand{\ra}{\rangle}
\newcommand{\la}{\langle}
\newcommand{\braket}[1]{\la#1 \ra}

\newcommand{\Eqref}[1]{Eq. \eqref{#1}}

\title{\Title}
\author[1,*]{Luca Leonforte}
\author[1,2]{Davide Valenti}
\author[1,3,4]{Bernardo Spagnolo}
\author[1,3]{Angelo Carollo}
\affil[1]{Department of Physics and Chemistry - Emilio Segré, Group of Interdisciplinary Theoretical Physics, University of
Palermo, Viale delle Scienze, Ed. 18, I-90128 Palermo, Italy}
\affil[2]{Istituto di Biomedicina ed Immunologia Molecolare (IBIM) "Alberto Monroy", CNR,
Via Ugo La Malfa 153, I-90146 Palermo, Italy}
\affil[3]{Radiophysics Department, Lobachevsky State University of
Nizhni Novgorod, 23 Gagarin Avenue, Nizhni Novgorod 603950, Russia}
\affil[4]{Istituto Nazionale di Fisica Nucleare, Sezione di Catania,
Via S. Sofia 64, I-95123 Catania, Italy} 

\affil[*]{luca.leonforte@unipa.it}

\begin{abstract}
We define the \emph{Uhlmann number} as an extension of the Chern number, and we use this quantity to describe the topology of 2D translational invariant 
Fermionic systems at finite temperature. We consider two paradigmatic systems and we study the changes in their topology through the \emph{Uhlmann number}.
Through the linear response theory we link two geometrical quantities of the system, the \emph{mean Uhlmann curvature} and the \emph{Uhlmann number}, to directly measurable physical quantities, i.e. the dynamical susceptibility and the dynamical conductivity, respectively. In particular, we derive a non-zero temperature generalisation of the Thouless-Kohmoto-Nightingale-den Nijs formula.
\end{abstract}

\begin{document}

\flushbottom \maketitle

\section*{Introduction}
The discovery of topological ordered phases (TOP) has attracted an
ever growing interest from the very outset~\cite{Bernevig2013},
partly due to the number of fascinating phenomena connected to it,
such as topologically protected edge
excitations~\cite{Hatsugai1993}, quantised current in insulating
systems~\cite{Klitzing1980,Thouless1982,Thouless1983,Niu1984,Nakajima2016,Tsui1982},
bulk excitations with exotic
statistics~\cite{Laughlin1983,Arovas1984,Nayak2008}. A relevant
subclass of TOP are the so called symmetry-protected TOP, which have
been extensively studied and classified thoroughly, according to a
set of topological
invariants~\cite{Altland1997,Schnyder2008,Ryu2010,Chiu2016}. The
above classification relies on the assumption that the relevant
features of a topological quantum
system are fundamentally captured by the system zero-temperature limit, i.e. by the properties of its pure ground state. 
However, the fate of these topological ordered phases
remains unclear, when a mixed state is the faithful description of
the quantum system, either because of thermal equilibrium, or due to
out-of-equilibrium
conditions~\cite{Magazzu2015,Magazzu2016,Guarcello2015,Spagnolo2015,Spagnolo2016,Spagnolo2018,Valenti2018,Spagnolo2018a}.
Over the last few years, different attempts have been done to
reconcile the above topological criteria with a mixed state
configuration~\cite{Avron2011,Bardyn2013,Huang2014,Viyuela2014,Viyuela2014a,Budich2015a,Linzner2016,Mera2017,Grusdt2017,Bardyn2018,He2018}.
The recent success of the Uhlmann approach~\cite{Uhlmann1986} in
describing the topology of 1D Fermionic
systems~\cite{Huang2014,Viyuela2014}, remains in higher
dimensions~\cite{Viyuela2014a} not as
straightforward~\cite{Budich2015a}. Moreover, the importance of this
approach and its relevance to directly observable physical
quantities still remains an interesting open question.

In this work, we propose to study 2D topological Fermionic systems,
at finite temperature, by means of a new set  of geometrical tools
derived from the Uhlmann approach~\cite{Uhlmann1986}, and more specifically from the \emph{mean
Uhlmann curvature} (MUC)~\cite{Carollo2018,Carollo2018a} . We study
2D-topological insulators (TIs), whose topological features are captured by the Chern
number~\cite{Chiu2016}. As with many other topological materials,
these systems may host gapless edge excitations, whose presence
characterises the onset of a non-trivial topological
phase~\cite{Laughlin1981}.

For translational invariant models, one can define the Chern number as
$\Ch = \frac{1}{2
\pi}\iint_{BZ} F_{x y}^B d^2 \mathbf{k}$,{\color{black} i.e. the integral over the Brillouin zone (BZ)} of the Berry curvature $F_{x y}^B$. The Ch is always an integer and it is the topological invariant that characterises the zero-temperature phase of the
system we are interested in. In order to study these models at finite temperature one should find a way to generalise the Chern number to a mixed state scenario. However, a
direct generalisation of the Chern number via the Uhlmann approach
leads to a trivial topological invariant. In this work, we construct
a quantity, \emph{the Uhlmann number} $n_U$, through the MUC. Stricktly speacking, this
quantity is not a topological invariant, but it provides a faithful description of topological and geometrical properties of the systems with respect to temperature changes. We apply these concepts to
two paradigmatic models of TI, the QWZ model\cite{Qi2006} and a TI
with high Chern number~\cite{Sticlet2012,Ivanov2001}, and explicitly derive the dependence of the
 Uhlmann number on temperature.  Beyond their
mathematical and conceptual appeal, we show that the MUC and
Uhlmann number are related to quantities directly accessible to
experiments, namely, the susceptibility to external perturbations
and the transverse conductivity.
\section*{Results}
\subsection*{Susceptibility and mean Uhlmann curvature}
The Uhlmann approach to geometric phase of mixed states allows to define a mixed state generalization of the Berry curvature, the \emph{mean Uhlmann curvature} (MUC).
The MUC can be defined as the Uhlmann geometrical phase over an infinitesimal loop (see section Methods )
\begin{equation}
\mathcal{U}_{\mu \nu} := \lim_{\delta_\mu \delta_\nu \rightarrow 0} \frac{\varphi^{U}[\gamma]}{\delta_\mu\delta_\nu }.
\end{equation}
The MUC is a geometrical quantity, whose definition relies on a rather formal definition of holonomies of density matrices. In spite of its abstract formalism, the MUC has interesting connections to a physically relevant object which is directly observable in experiments, the susceptibility.
By using the linear response theory, we can indeed relate the MUC to the dissipative part of the dynamical susceptibility. Indeed, one can consider the most general scenario of a system with a Hamiltonian $\HH_0$, perturbed as follows
\begin{equation}
\label{eq:pert} 
\HH = \HH_0 + \sum_\mu \hat{O}_\mu \lambda_{\mu},
\end{equation}
where $\{\hat{O}_{\mu}\}$ is a set of observables of the system, and
$\{\lambda_{\mu}\}$ is the corresponding set of perturbation
parameters.
Then, we show (see section Methods) that for a thermal state, the dissipative part of the dynamical susceptibility $\chi_{\mu \nu}''(\omega,\beta)$ is related to the MUC as follows 
\begin{equation}
\label{ulmsuc}
\mathcal{U}_{\mu \nu} =  \frac{i}{  \pi \hbar }  \int_{-\infty}^{+\infty}   \frac{d\omega}{\omega^2} \tanh^2 \left( \frac{\omega \beta}{2} \right)\chi_{\mu \nu}''(\omega,\beta) ,
\end{equation}
where the set of perturbations $\{\lambda_\mu\}$ in~(\ref{eq:pert}) plays the role of the parameters in the derivation of $\mathcal{U}_{\mu \nu}$ {\color{black}, and where $\beta:=1/k_B T$, is the inverse of the temperature}.
Moreover, by means of the fluctuation-dissipation theorem~\cite{Altland2006}, one can also derive a further expression for~\Eqref{ulmsuc} in terms of the dynamical structure factor, 
$S_{\mu \nu}(\omega,\beta) = \int_{-\infty}^{+\infty} d t e^{i \omega t} S_{\mu \nu} (t)$, (i.e. the Fourier transform of the correlation matrix $S_{\mu \nu}(t) = \braket{\hat{O}_\mu (t) \hat{O}_\nu (0)}$) 
namely
\begin{equation}
\label{MUCCorr1}
\mathcal{U}_{\mu \nu} =  \frac{i}{2 \pi \hbar }  \int_{-\infty}^{+\infty} \frac{d\omega}{\omega^2} \tanh^2 \left( \frac{\omega \beta}{2} \right) ( S_{\mu \nu}(\omega,\beta) - S_{\nu \mu}(-\omega) ).
\end{equation}
Equations~(\ref{ulmsuc}) and~(\ref{MUCCorr1}) provide a means to explore experimentally the geometrical properties of physical systems via the dissipative part of the dynamical susceptibility, and the imaginary part of the (off-diagonal)-dynamical structure factor.   

Beyond its geometrical meaning, one can also show that the MUC has profound interpretation in terms of quantum multi-parameter estimation theory~\cite{Ragy2016,Nichols2018,Carollo2018,Carollo2018a,Safranek2018}. Indeed, the uncertainty in the estimation 
of a set of parameters $\{\lambda_{\mu}\}$ of a physical system is lower bounded by the Cramer-Rao (CR) bound~\cite{Holevo2011,Paris2009,Helstrom1976}, i.e. $\text{Cov}(\hat{\lambda})\geq J^{-1}$, 
where $J$ is the quantum Fisher information matrix, whose elements are $\munu{J}=\frac{1}{2}\Tr[\rho \{L_\mu, L_\nu \}]$, and $\text{Cov}(\hat{\lambda})$ is the covariance matrix, 
which quantifies the uncertainty on $\{\lambda_{\mu}\}$. Both in a \emph{classical multi-parameter} and in a \emph{quantum single-parameter} estimation problem, the CR bound is always tight. 
However, in the \emph{quantum multi-parameter} case, the CR bound may not be saturated, due to a manifestation of the uncertainty principle, known as \emph{incompatibility condition}~\cite{Ragy2016,Nichols2018,Safranek2018}.
Such an incompatibility is quantified by the MUC~\cite{Ragy2016,Carollo2018}, which signals whether the estimation of a set of parameters is hindered by the inherent quantum nature of the 
underlying physical system.

Thanks to \Eqref{ulmsuc} we see that if the perturbations are longitudinal, so that they affect only the expectation value of the correspondent operator, then the MUC must be zero, and so the two parameters are \emph{compatible}. On the converse, a transverse susceptibility signals the presence of an incompatibility which emerges from to the quantum nature of the physical system.
\subsection*{Electrical conductivity and $\bm{n_U}$}
The geometrical interpretations of the MUC as a generalisation of the Berry curvature and its connection to physically accessible quantities are quite desirable features. One may wonder whether these properties may be used to construct a physically appealing finite-temperature generalisation of a topological invariant, i.e. the Chern number. 

The Chern number, $ \Ch = \frac{1}{2\pi} \int_{BZ} F_{x y}^{B} d k_x d k_y$, is the invariant that characterises the topology of the bands in 2D translational invariant systems, where  $F_{xy}^{B}$ {\color{black} is the Berry curvature}.
A natural finite temperature generalisation of the $\Ch$ can be constructed out of MUC, $\munu{\mathcal{U}}(\mathbf{k})$ (see section Methods), as
\begin{equation} 
\label{nu1} 
n_U = \frac{1}{2\pi} \int_{BZ} \munu{\mathcal{U}} d k_\mu d k_\nu.
\end{equation}
$n_{U}$ is clearly a finite temperature generalisation of the Chern number, to which it converges in zero temperature limit. One should notice, however, that $n_U$ is not itself a topological invariant, as it is not always an integer. Nevertheless, it provides a measure of the geometrical properties of the system and, above all, $n_{U}$ posses quite remarkable connections to quantities which are readily accessible in experiments.

Indeed, consider a translational invariant 2D Fermionic system. In the quasi-momentum representation, the Hamiltonian reads $\HH_{0} = \sum_{\mathbf{k}\in BZ} \funk{\HH}$. 
When the system is perturbed by a time-dependent homogeneous electric field, one can show that the dissipative part of the dynamical transversal conductivity is directly linked to the Uhlmann number (\Eqref{nu1}) via the following expression (see section Methods)
\begin{equation} \label{tknnfor1} \frac{1}{\pi}\int_{-\infty}^{+\infty}
\frac{d\omega}{\omega}\tanh^2 \left( \frac{\hbar \omega \beta}{2}
\right)  \sigma''_{x y} (\omega,\beta) = - \frac{e^2}{2 \pi \hbar} n_U .
\end{equation}
{\color{black} From the definition and the properties of $ \sigma''_{\mu \nu} (\omega,\beta) $ (see section Methods), \Eqref{tknnfor1} can be rewritten as 
\begin{equation}\label{nUvsSigmaR}
n_U \frac{e^2}{2 \pi \hbar} = -\int_{-\infty}^{+\infty} d \omega \tilde{\sigma}_{xy}(\omega,\beta) K_\beta(\omega),
\end{equation}
where $\tilde{\sigma}_{xy}(\omega,\beta):=\text{Re}[\sigma_{xy}(\omega,\beta)-\sigma_{yx}(\omega,\beta)]/2$ is the real, antisymmetric part of the transverse conductivity, and the kernel $K_\beta(\omega)$ is a probability density function over the frequency domain $\omega\in\mathbb{R}$, that tends to the Dirac $\delta(\omega)$ in the zero temperature limit.
The expression in~\Eqref{nUvsSigmaR} is clearly a finite-temperature extension of the famous Thouless-Kohmoto-Nightingale-den Nijs (TKNN) formula~\cite{Thouless1982} , i.e.}
\begin{equation}\label{eq:tknn} \sigma_{xy} = -\Ch \frac{e^2}{h},
\end{equation}
which connects the transversal conductivity of a topological insulator to the Chern number. In the same spirit, \Eqref{tknnfor1} {\color{black} and ~\Eqref{nUvsSigmaR} provide} a relation, valid at any temperatures, between the transversal conductivity and the geometrical properties of the band structure described by $n_U$. {\color{black} A relevant difference between Eqs~\eqref{nUvsSigmaR} and~\eqref{eq:tknn} is that the latter involves an average of the dynamical conductivities on a frequency band peaked around $\omega=0$, with a width $\Delta \omega\propto 1/\hbar\beta$. Nevertheless, Eqs.~\eqref{tknnfor1} and~\eqref{nUvsSigmaR} provide the operational means to probe experimentally the geometrical properties of the system at any finite temperature.}  

Moreover, combining \Eqref{ulmsuc} and \Eqref{tknnfor1} we get {\color{black} 
\begin{equation} \label{nuMUCE} \mathcal{U}_{E_x E_y} =
-\frac{e^2}{\hbar^2} 2\pi n_U ,
\end{equation}
 where $\mathcal{U}_{E_x E_y}$ is the MUC, in which, two orthogonal components $E_x$ and $E_y$ of the  electric field take the role of the parameters $\{\lambda_\mu\}$ with respect to which the MUC is calculated. 
Hence, equation~\eqref{nuMUCE} links the topology
of the system to the MUC~(see section Methods), derived with respect
to physically accessible external parameters, namely the electric fields. Interestigly, one can also show~\cite{Ragy2016,Carollo2018a,Carollo2018} that the MUC has a very profound interpretation in terms of quantum
estimation theory. Namely, $\mathcal{U}_{\mu\nu}$ marks the
\emph{incompatibility} of two parameters $\lambda_\mu$ and $\lambda_\nu$, in the sense specified in~\cite{Ragy2016,Carollo2018a,Carollo2018}, when these parameters needs to be evaluated simultaneously by any quantum multi-parameter estimation protocol. This incompatibility is a manifestation of the quantum uncertainty-principle, arising from the inherent quantum nature of the underlying physical
system. When applied to~\Eqref{nuMUCE}, this argument links the presence of a non-trivial topology in the system to an \emph{incompatibility} between the orthogonal components $E_x$ and $E_y$ of the electric field, in a quantum estimation protocol.
}

In the following two subsections we will apply some of the general considerations described so far to two archetypical models of 2D topological insulator.
\subsection*{A two-dimensional topological insulator with high Chern number}
\begin{figure}[t!]
	\includegraphics[width = \linewidth ]{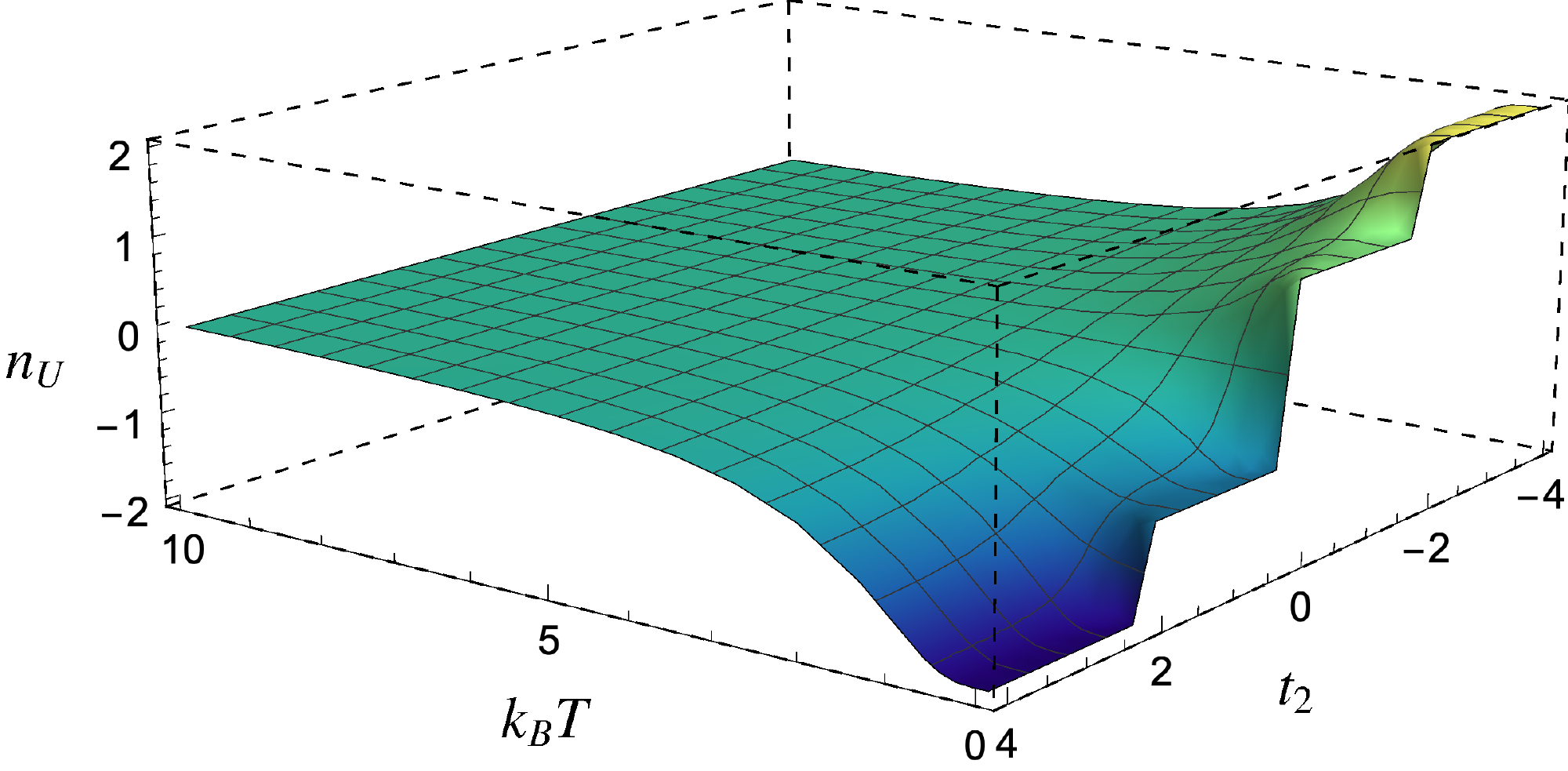}
	\caption{The graph shows how $n_U$ changes for a topological
		insulator, with high Chern number, as a function of the temperature
		and the hopping term $t_2$.} \label{fig:HighChern3D}
\end{figure}

A prototypical example of a 2D Chern insulator is a model that was first proposed by D. Sticlet et
al.~\cite{Sticlet2012}. This is a topological insulator of Fermions lying on the vertices of a triangular lattice. Each Fermion carry a two-dimensional
internal degree of freedom. By tweaking the interaction parameters, this model can be tuned to up to
five different topological phases. Here, we consider Sticlet's model with the following parametrisation 
\begin{equation}
\HH \!\!=\!\! \sum_{ij} \Big[  c_{i+1,j}^{\dagger} (t_1 \sigma_1 + i t_3 \sigma_3) c_{i,j} + c_{i,j+1}^{\dagger} ( t_1 \sigma_2 + i t_3 \sigma_3 ) c_{i,j} 
 + c_{i+1,j+1}^{\dagger} t_2 \sigma_3 c_{i,j} + \text{H.c.} \Big].
\end{equation}
{\color{black} The Pauli matrices describe the internal degree of freedom and $t_i$ is a hopping amplitude coupling nearest neighbour Fermions 
with different orbitals.} In the momentum representation the Hamiltonian reads
\begin{equation}
\funk{H} = 2 \{\cos (k_x)\sigma_1 + \cos (k_y)\sigma_2
+ [ t_2 \cos ( k_x + k_y ) + \sin (k_x) + \sin (k_y) ]\sigma_3 \},
\end{equation}
where we have set $t_1 = t_3 =t=1$, and all the energies are scaled with respect to these parameters. The topological phases at zero
temperature are characterised by the Chern number, {\color{black} whose value, as a function of $t_2$, reads as}
\begin{equation} \Ch = \begin{cases}
+2 , \qquad \text{ if  }t_2 < -2 \\
+1  , \qquad \text{ if  } -2 < t_2 < 0 \\
-1 , \qquad  \text{ if  } 0 < t_2 < 2 \\
-2 , \qquad \text{ if  }t_2 > 2 .
\end{cases}
\end{equation}
Notice that this model carries a non-trivial zero-temperature topological phase (i.e. $\Ch \neq 0$) for the whole parameter space. We consider the system in a thermal Gibbs state and we numerically calculate the Uhlmann number (see~\Eqref{nu}), whose values are graphically represented as a
function of $t_2$ and temperature $T$ in
Fig.~\ref{fig:HighChern3D}.
\begin{figure}[!h]
	\centering
	\includegraphics[width = 1.05\linewidth ]{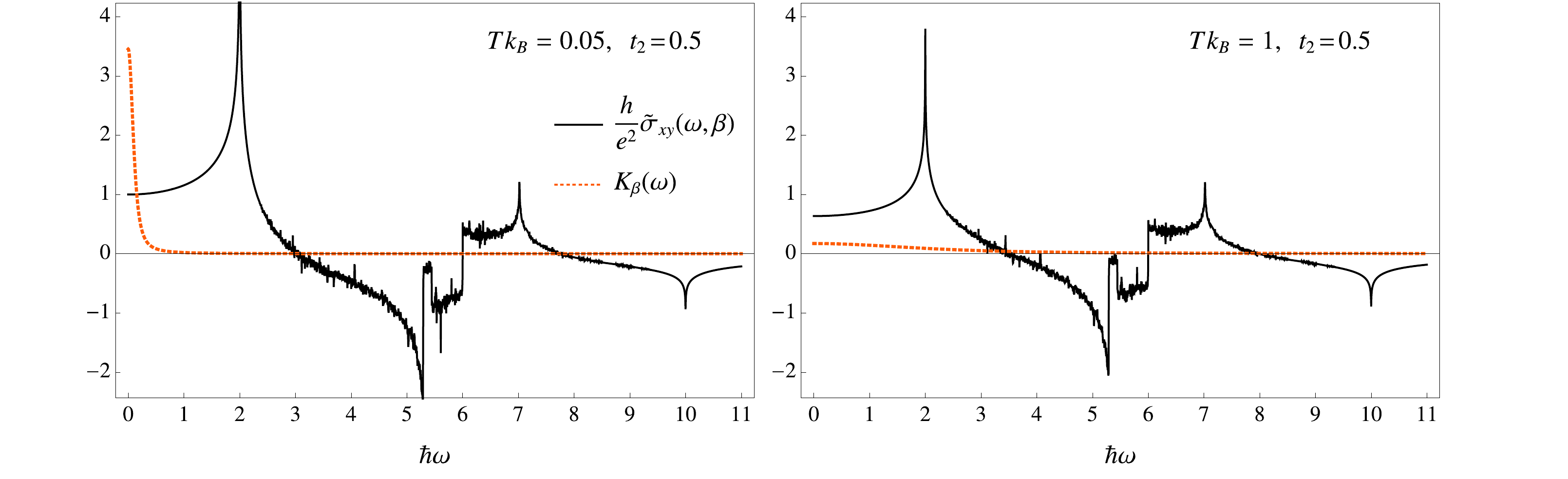}
	\caption{\color{black}  The graphs display the dependence of $\tilde{\sigma}_{xy}(\omega)$ (black, solid line) and the Kernel $K_\beta(\omega,\beta)$ (orange, dashed line), on the frequency $\omega$ for $T k_B=0.05$ and $T k_B=1$ and parameter $t_2=0.5$. The transverse conductivity $\tilde{\sigma}_{xy}(\omega)$ displays van~Hove singularities across the single particle spectrum of the model, which, for $t_2=0.5$, ranges from $\omega=2$ to $\omega=10$. $K_\beta(\omega)$ is centered around $\omega=0$ and approximately non-vanishing only below the frequency bandwidth of $\Delta \omega \simeq \frac{10}{\hbar \beta}$.}\label{fig:CondHC}
\end{figure}
As expected, the $n_U$ correctly describes the
topological phase transition at zero temperature. For high
temperatures, the behaviour of $n_U$ shows a typical cross-over
transition, without any criticality between different regions~\cite{Viyuela2014,Budich2015a,Mera2017}.
One can observe a smooth monotonic vanishing of $n_{U}$ as the temperature
increases.

{\color{black}
In order to grasp a better understanding of the relation, predicted by Eq.~\eqref{nUvsSigmaR}, between $n_U$ and  the real conductivity, we consider the behaviour of $\tilde{\sigma}_{x y}$ and $K_\beta$ with respect to frequency and temperature. Fig~\eqref{fig:CondHC} graphically shows $\tilde{\sigma}_{xy}$ and the probability density function $K_\beta$ as a function of $\omega$ for two temperatures, $T k_B=0.1$ and $T k_B=2$, and for $t_2=0.5$ (corresponding to a zero-temperature Ch$=-1$). As expected, for small temperatures the real transverse conductivity approaches the value $\tilde{\sigma}_{x y}(0)\frac{\hbar}{e^2}\simeq-\textrm{Ch}=1$. The figure shows the distinctive dependence of the conductivity on the density of states (see~\Eqref{eq:TCond}), featuring van~Hove singularities across the single particle frequency band. The latter, for the chosen parameter $t_2=0.5$, extends from $\omega=2$ to $\omega=10$. For the same values of the parameters, the shape of the probability density function $K_\beta$ shows strong dependence on temperature. The distribution is sharply peaked around the static conductivity for small values of temperature, and broadens up for higher values of $T$. This explains, on the one hand, the strong dependence of $n_U$ on temperature, and, on the other hand, the rather weak dependence of $n_U$ on the dynamical conductivity even for relatively small values of the frequencies. As a consequence, the singular features of $\tilde{\sigma}_{xy}$ are not observable in $n_U$, because they are either neglected by $K_\beta$, for small values of $T$, or washed out in the averaging process, as $T$ grows.
}
\subsection*{QWZ model}
In this section we consider the QWZ model, introduced by Qi, Wu and Zhang \cite{Qi2006,Asboth2016} as an archetypical example of topological insulator. The QWZ Hamiltonian is constructed from the Rice-Mele model, where time is promoted to a spatial dimension. This system provides the simplest example of an anomalous quantum Hall system. 
The QWZ is a model of Fermions on a square lattice, with a two-dimensional orbital degrees of freedom per site, and its Hamiltonian is given by
\begin{equation}
\HH = J\sum_{ij} \left[c^{\dagger}_{i+1,j} \left( \frac{\sigma_z + i \sigma_x }{2}\right) c_{i,j} + c^{\dagger}_{i,j+1} \left( \frac{\sigma_z + i \sigma_y }{2}\right) c_{i,j} + H.c. \right] +  u J \sum_{ij} c^{\dagger}_{i,j} \sigma_z c^{\dagger}_{i,j} ,
\end{equation}
%
where $\sigma_i$ are the Pauli matrix and $J$ fixes the global energy scale, and for simplicity we set $J=1$.
The single-particle Hamiltonian in the quasi-momentum representation is
\begin{equation} \funk{H} = \{\sin{k_x}
\sigma_x + \sin{k_y}\sigma_y + [u + \cos{k_x} + \cos{k_y}]
\sigma_z\},
\end{equation}
where the $\sigma_{i}$ act on the orbital degrees of freedom. The topological
phases of the model at $T=0$ are characterised by the following
Chern numbers as a function of $u$
\begin{equation} \Ch = \begin{cases}
0 , \qquad \text{ if  } u < -2 \\
1  , \qquad \text{ if  } -2 < u < 0 \\
-1  , \qquad \text{ if  } 0 < u < 2 \\
0 , \qquad \text{ if  } u > 2.
\end{cases}
\end{equation}
\begin{figure}[t!]
\centering
\includegraphics[width=\linewidth]{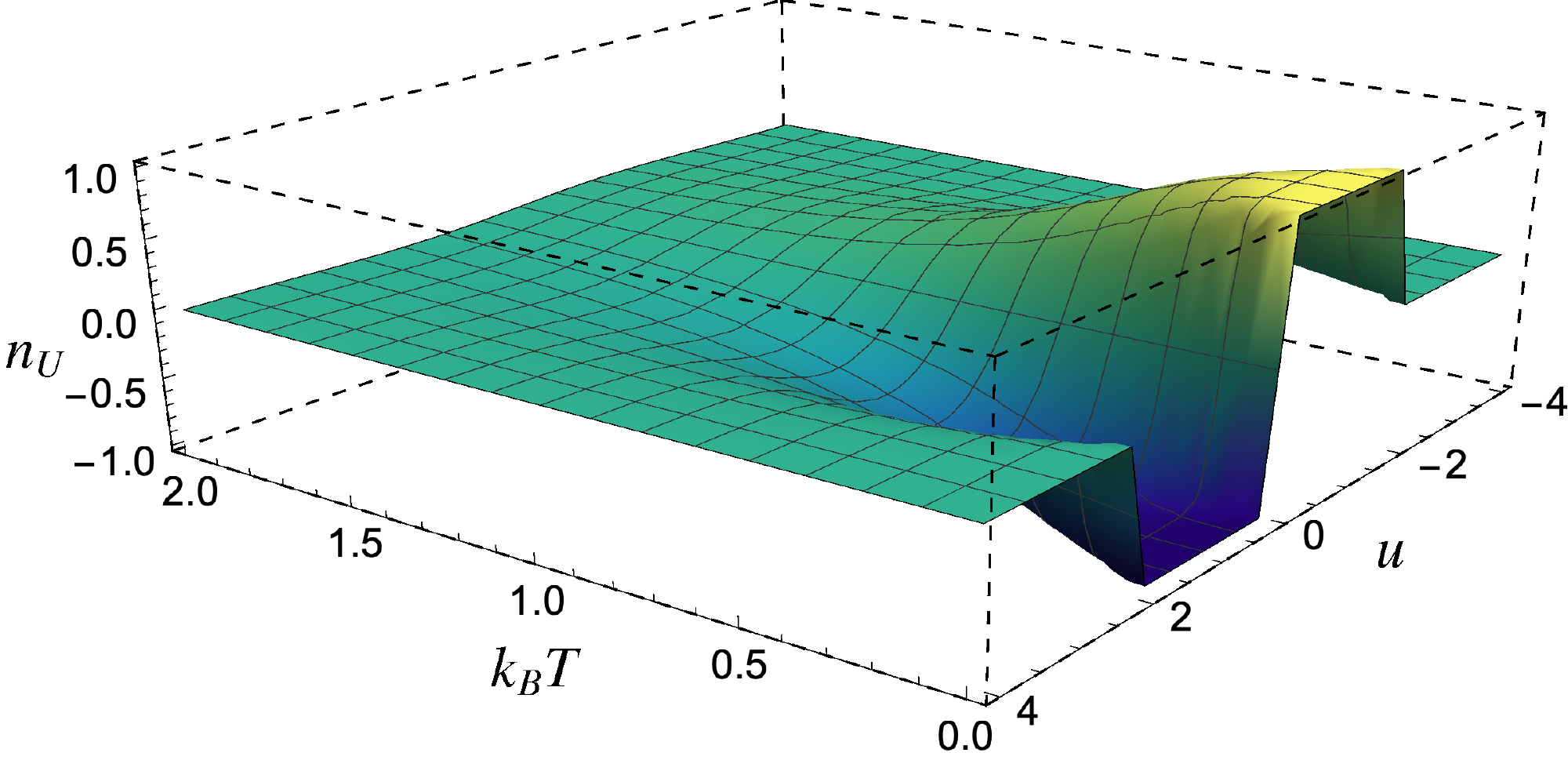}
\caption{\color{black} QWZ model: Uhlmann number behaviour as a function
of temperature $T$ and the parameter u.} \label{Pw3d}
\end{figure}
\begin{figure}[t!]
\centering
 \includegraphics[width=\linewidth]{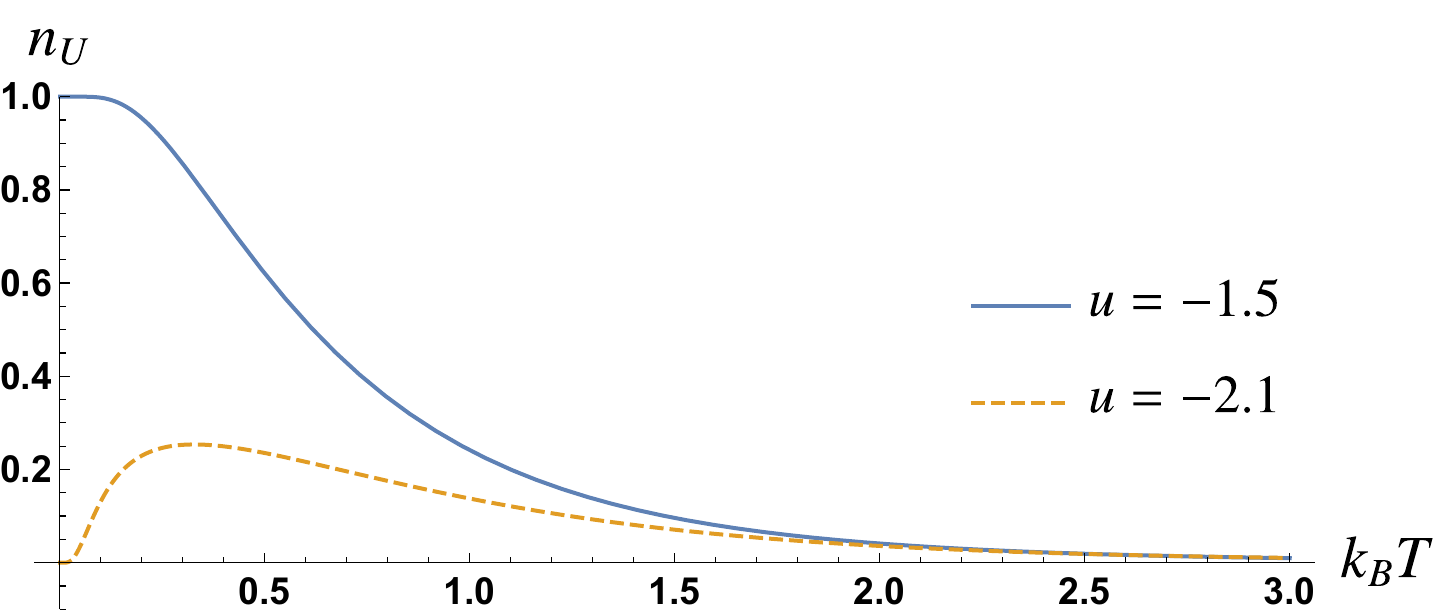}
\caption{\color{black} QWZ model: Uhlmann number behaviour as a function of temperature $T$ for two different values of the parameter $u$, namely $u = -1.5$ and $u = -2.1$.} \label{fig:Pw2d}
\end{figure}
\begin{figure}[t!]
	\centering
	\includegraphics[width = 1.05\linewidth ]{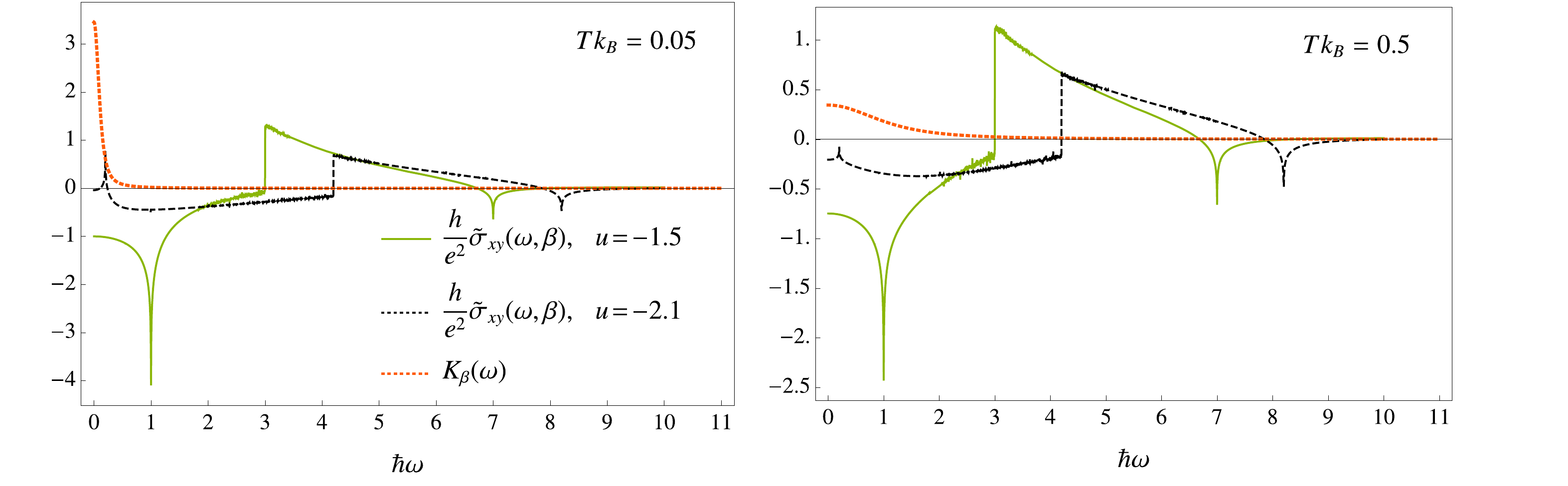}
	\caption{\color{black} The graphs plots the dependence of real transverse conductivity  $\tilde{\sigma}_{xy}(\omega,\beta)$ (in units of $e^2/h$) and the Kernel $K_\beta(\omega)$, on the frequency $\omega$ for two temperatures, $Tk_B=0.05$ and $Tk_B=0.5$. One can appreciate in both plots the presence of van Hove singularities. {\color{black} In particular, one can observe the appearance of a singularity at $\omega=\Delta$, i.e. the band-gap of the model, which is $\hbar\Delta=1$ for $u=-1.5$ and $\hbar\Delta=0.2$ for $u=-2.1$. For $u=-2.1$, the presence of a singularity so close to $\omega=0$ accounts} for the non-monotonic behaviour shown in Fig.~\ref{fig:Pw2d}, displayed by $n_U$ as $T$ increases.} \label{fig:CondQWZ}
\end{figure}
For topological non-trivial regions, $\Ch = \pm 1$, the system
presents chiral edges states, as in the integral quantum Hall effect.
We assume a thermal Gibbs state, and numerically calculate the Uhlmann
number (see \Eqref{nu}), whose values are graphically represented in
Fig.~\ref{Pw3d}. As expected, the $n_U$ correctly describes the
topological phase transition at zero temperature. For high
temperatures, the behaviour of $n_U$ shows a typical cross-over
transition, without any criticality between different regions.
One can observe a smooth vanishing of $n_{U}$ as the temperature
increases.

By fixing $u$ in a specified phase, one can see two different
dependencies of $n_U$ as temperature increases. In a non-trivial
topological phase, e.g. when $\Ch =\pm 1$, we see that $n_U$
vanishes monotonically (see the blue solid line in
Fig.~\ref{fig:Pw2d} ). On the other hand, one can see a peculiar
non-monotonic behaviour of $n_U$ in the trivial phase, for
values of the parameter $u$ in the close proximity of the critical point (see
the dashed orange line in Fig.~\ref{fig:Pw2d}).

This can be interpreted as a thermal activation of the topological
property of the system. Indeed, in a phase, which is trivial at zero
temperature, there may be a range of temperatures for which the
geometrical properties of the bands show non-trivial values. This
can be explained by a thermal transfer of population from the 
valence to the conduction band, in the regions of the Brillouin zone 
in which the gap is smaller. These are the regions which contribute the most
to the Uhlmann curvature, overall providing a non-trivial net value of the
Uhlmann number. The closer the system is to a critical point, (for
example for $u\to -2^{-}$ in the QWZ model), the more
pronounced this effect is. {\color{black} This is due, on the one hand, by the narrowness of the
gap which allows the valence band in this region of the BZ to be populated for relatively small values of $T$, and on the other hand, by the nearly divergent behaviour of the Berry curvature in the
vicinity of the gap.}

{\color{black} In Fig.~\ref{fig:CondQWZ} we plot the dependence of $K_\beta$ (orange dotted line) and $\tilde{\sigma}_{xy}$ on frequency for two values of temperature $T k_B=0.1$ and $T k_B=0.9$ and for the two different values of the parameter $u$ considered in Fig.~\ref{fig:Pw2d}. For $u=-1.5$ (green solid line) the model is in a topological phase at zero temperature (Ch$=1$), while for $u=-2.1$ (black dashed line) the system is in a trivial zero-temperature phase (Ch$=0$), but in close proximity to the critical value $u=-2$. As for the previous model, considered in Fig.~\ref{fig:CondHC}, one can observe the appearance of van Hove singularities in transverse conductivity. {\color{black} Interestingly, for $u=-2.1$, one can observe the singularity at $\hbar\omega=0.2$, corresponding to the band gap of the model, which for $u=-2.1$ is given by $\hbar\Delta=0.2$. Clearly, as the model becomes critical, at $u\to -2$, this peak will shift towards $\omega=\Delta \to 0$. The presence of such a singularity for small values of $\omega$} explains the non-monotonic behaviour displayed by $n_U$ in Fig.~\ref{fig:Pw2d}. For $T<<1$, the distribution $K_\beta$ is strongly peaked at $\omega=0$, and only the (trivial) static conductivity contributes to $n_U$. As $T$ increases, $K_\beta$ broadens up, and picks up non-trivial contributions, mostly due to the singularity at $\omega=\Delta$. 

This explanation of the non-monotonicity of $n_U$'s behaviour is consistent with the interpretation in terms of thermal activation of the topological properties of the system. By considering formula for $\tilde{\sigma}_{xy}$ (see Eq.~\eqref{eq:TCond} in Methods), one realises that the peak of $\tilde{\sigma}_{xy}$ at the singular value $\hbar\Delta=0.2$ carries information on the Berry curvature $F^{B}_{xy}$ in the Brillouin zone around at the band gap $\hbar\Delta$. Close to criticality, this is the region that contributes the most to the overall value of zero-temperature Chern-number.
}
\section*{Discussion}
We have introduced the concept of \emph{Uhlmann number}~(see
Methods), as a finite temperature generalisation of the Chern
number. Beyond its mathematical and conceptual appeal, we have
linked the \emph{Uhlmann number} to directly measurable physical
quantities, such as the dynamical susceptibility~(see section Methods) and
dynamical structure factor. We have shown that, in 2D translational
invariant Fermionic systems, the above quantities can be
straightforwardly measured through dynamical conductivity. This
leads to a connection between Uhlmann number and transversal
conductivity that may be thought as a finite-temperature
generalisation of the (TKNN) formula. Moreover, these expressions
highlights also a relation between the MUC, in the electric field
parameters space, and $n_U$. The latter shows that a non-trivial
topology gives rise to an \emph{incompatibility} condition in the
parameter estimation problem of two orthogonal components of the
electric field, due to the inherent quantum nature of the underlying
physical system.

\section*{Methods}
\subsection*{The Uhlmann number}
The Uhlmann Geometric Phase is a generalisation of the Berry phase
when the system is in a mixed state~\cite{Uhlmann1986}. This
generalisation relies on the idea of \emph{amplitude} of a density
operator $\rho\in{\mathcal{B}}(\HH)$, which is defined as an
operator $\omega$ satisfying $\rho= \omega \daga{\omega}$. Such a
definition leaves a $U(n)$ gauge freedom on the choice of $\omega$,
as any operator $\omega' = \omega U$, with $U$ unitary matrix,
fullfils the same condition $\rho= \omega' \daga{\omega'}$. Let
$\rho_\lambda$ be a family of density matrices parametrized by
$\lambda \in \mathcal{M}$, with $\gamma := \{ \lambda(t) \in
\mathcal{M}, t \in [0,T] \}$ a smooth closed curve in a parameter
manifold $\mathcal{M}$ and $\omega_\lambda$ the corresponding path
of amplitudes. To reduce the gauge freedom, Uhlmann introduced a
parallel transport condition on $\omega_\lambda$~\cite{Uhlmann1986}.
When this condition is fulfilled on a closed curve $\gamma$, the
amplitudes at the endpoints of the curve must coincide up to a
unitary transformation $\omega_{\lambda(T)} = \omega_{\lambda(0)}
V_\gamma$, where $V_\gamma$ is the holonomy associated to the
path~\cite{Uhlmann1986}.

The holonomy is expressed as $V_\gamma = \mathcal{P} e^{i \oint A}$,
where $\mathcal{P}$ is the path ordering operator, and $A = \sum_\mu
A_\mu d \lambda_\mu $ is the Uhlmann connection one-form, the
non-Abelian generalization of the Berry connection. The Uhlmann
connection is defined by the following
ansatz~\cite{Uhlmann1989,Dittmann1999} $ \partial_\mu \omega =
\frac{1}{2} L_\mu \omega - i \omega A_\mu $, where $L_\mu$ are the
Hermitian operators known as symmetric logarithmic derivative (SLD),
and $(\partial_\mu = \partial / \partial \lambda_\mu )$ is the
derivative with respect to a parameter in the manifold
$\mathcal{M}$. The SLD is defined as the operator solution of the
equation $\partial_\mu \rho = \frac{1}{2} \{ L_\mu , \rho \}$. The
components of the Uhlmann curvature, the analogue of the Berry
curvature, are defined as $F_{\mu \nu} = \partial_{\mu} A_\nu -
\partial_\nu A_\mu - i [A_\mu , A_\nu]$. They can be understood in
terms of the Uhlmann holonomy per unit area associated to an
infinitesimal loop, $F_{\mu \nu} = \lim_{\delta_\mu \delta_\nu
\rightarrow 0} i \frac{1 - V_\gamma}{\delta_\mu \delta_\nu}$, where
$\delta_\mu \delta_\nu$ is the area of the infinitesimal
parallelogram spanned by the two independents direction $\delta_\mu
\hat{e}_\mu$ and $\delta_\nu \hat{e}_\nu$.

The Uhlmann phase is defined as $\varphi^{U}[\gamma]  = \arg  \Tr
[\daga{\omega_{\lambda(0)}}\omega_{\lambda(T)}]$. The mean Uhlmann
curvature~\cite{Carollo2018}, defined as the Uhlmann phase per unit
area for an infinitesimal loop, is given by
\begin{equation} 
\mathcal{U}_{\mu \nu} := \lim_{\delta_\mu \delta_\nu \rightarrow 0} \frac{\varphi^{U}[\gamma]}{\delta_\mu
\delta_\nu } = \Tr [ \daga{\omega_{\lambda(0)}} \omega_{\lambda(0)}
F_{\mu \nu}].
\end{equation}
One can show that the MUC can be expressed in terms of the SLD in a very convenient way as
\begin{equation} \label{MUC} \mathcal{U}_{\mu \nu} =
\frac{i}{4} \Tr [ \rho [L_\mu , L_\nu ] ].
\end{equation}
 One can easy show that the MUC converges, in the pure state limit, to the Berry curvature
$F_{\mu \nu}^{B}$.

The systems we study in this work are 2D translational invariant
systems whose topology is characterised by the Chern number of the
ground state, that is
\begin{equation} \label{Chern} \Ch = \frac{1}{2\pi} \int_{BZ}
F_{x y}^{B} d k_x d k_y , 
\end{equation}
i.e. the integral over the first Brillouin zone (BZ) of the Berry
curvature $F_{xy}^{B} = \partial_x A_y^B - \partial_y
A_x^B$,  where $A_\mu^B = i \braket{\psi_k | \partial_\mu |
\psi_k}$ is the Berry connection of the ground state. Here the
parameter manifold is the BZ itself, i.e. $\partial_\mu = \partial /
\partial k_\mu $, with $\mu,\nu \in \{x,y\}$.

Similarly, one can define the following quantity, the \emph{Uhlmann
number}, as the integral over the BZ of the MUC
\begin{equation} \label{nu} n_U = \frac{1}{2\pi} \int_{BZ}
\mathcal{U}_{xy} d k_x d k_y,
\end{equation}
where, in analogy with~eq.\eqref{Chern}, $\mathcal{U}_{xy}$ is the MUC of Eq.~\eqref{MUC}, where the parameters $\{\lambda_{\mu}\}$ are identified with the quasi-momenta $k_x$ and $k_y$. 
$n_{U}$ is clearly a finite temperature generalisation of the Chern
number, to which it converges in zero temperature limit. One easily
sees that the MUC, and hence the $n_U$, is gauge invariant, i.e. it
does not depend on the gauge choice of the amplitude. Nonetheless
$n_U$ is not a topological invariant, and it is not always an
integer as the Chern number is. In this work, we use $n_U$ as
an extension of the Chern number and we will link this quantity to
physical proprieties of the systems.

In order to do this, let's consider a 2D translational
invariant systems, which may show non-trivial topology at zero
temperature. The Hamiltonian of these systems can be cast in the following form, {\color{black}
\begin{align}\label{eq:k-H}
	\HH = \sum_{\mathbf{k}\in BZ} \pedk{\daga{\Psi}}\funk{H} \pedk{\Psi},
\end{align}
 where the first quantized Hamiltonian $\funk{H}$, for two-band systems,} is a $2\times2$ matrix.
The latter can be written as $\funk{H} = \varepsilon_{\mathbf{k}}
\mathds{1} + \vec{h}_\mathbf{k} \cdot \vec{\sigma}$, where the 
$\vec{h}_\mathbf{k}$ is a 3D vector and $\vec{\sigma}$ are the
Pauli matrices. $\pedk{\Psi}$ {\color{black} are Nambu spinors, which for two-band topological insulators} are
$\pedk{\Psi} := (a_\mathbf{k}, b_\mathbf{k})^t$, with $a_\mathbf{k}$
and $b_\mathbf{k}$ Fermionic annihilation operators of two different
species of Fermions of the system. {\color{black} The Berry curvature assumes the
following form,
\begin{equation} \label{FtwoD} {F}^B_{xy} = \frac{1}{2}
(\partial_{x}\hat{h}_\mathbf{k}\times
\partial_{y}\hat{h}_\mathbf{k}) \cdot \hat{h}_\mathbf{k},
\end{equation}}
where $\hat{h}_\mathbf{k} = \vec{h}_\mathbf{k} /
|\vec{h}_\mathbf{k}|$.

At thermal equilibrium, {\color{black} i.e. assuming a Gibbs state} $\rho = \frac{e^{-\beta
\HH}}{\mathcal{Z}} $, where $\beta = 1/k_b T$ is the inverse of the temperature and $\mathcal{Z} = \Tr [ e^{-\beta
\HH} ]$ is the partition function, {\color{black} the MUC $\mathcal{U}_{xy}$, calculated form Eq.~\eqref{MUC} with respect to the parameters $k_x$ and $k_y$, reduces to the following simple expression
\begin{equation} \label{MUCtwoD}
\mathcal{U}_{xy} = \tanh\left(\frac{\beta
|\vec{h}_\mathbf{k}|}{2}\right)\tanh^2  (\beta |\vec{h}_\mathbf{k}|) \cdot
F^B_{xy}.
\end{equation}
In this form the MUC appears as a straightforward modification of the Berry curvature $F^B_{xy}$, to which it manifestly converges in the $\beta\to\infty$ limit.
}
\subsection*{Susceptibility and MUC}
By using the linear response theory, we now derive a remarkable relation between the MUC, an inherent geometrical quantity, to a physically relevant quantity, the susceptibility. Let's consider a system with a Hamiltonian $\HH_0$, perturbed as follows
\begin{equation}
\label{eq:pert1} \HH = \HH_0 + \sum_\mu
\hat{O}_\mu \lambda_{\mu},
\end{equation}
where $\{\hat{O}_{\mu}\}$ is a set of observables of the system, and $\{\lambda_{\mu}\}$ the corresponding set of sources. 
We are considering the system in thermal equilibrium, i.e. $\rho = \frac{e^{-\beta \HH}}{\mathcal{Z}} $, where $\mathcal{Z} = \Tr [ e^{-\beta \HH} ]$ is the partition function. 
The dissipative part of the dynamical susceptibility, with respect to $\hat{O}_\mu$ is defined as:
\begin{equation}
\chi''_{\mu \nu} (t) = \frac{1}{2 \hbar} \braket{ [ \hat{O}_{\mu} (t), \hat{O}_{\nu} ] }_0
\end{equation}
One can show that the Fourier transform of the dissipative part of the dynamical susceptibility has the following expression in the Lehmann representation
\begin{equation} 
\chi_{\mu \nu}''(\omega,\beta)= \frac{\pi}{\hbar}\sum_{i j} (\hat{O}_{\mu})_{i j} (\hat{O}_{\nu})_{j i} (p_i - p_j) \delta (\omega + \frac{E_i - E_j}{\hbar}),
\end{equation}
{\color{black} where $p_i$'s are the eigenvalues of the density matrix in the Boltzmann-Gibbs ensemble, i.e. $p_i=e^{-\beta E_i}/Z$, and $E_i$'s are the corresponding Hamiltonian eigenvalues. For thermal states, one can exploit} the identity $\frac{p_i- p_j}{p_i + p_j}= \int_{-\infty}^{+\infty} d\omega \tanh \left( \frac{\hbar \omega \beta}{2} \right) \delta (\omega + \frac{E_i - E_j}{\hbar})$, {\color{black} which leads to the following relation between} the $\chi_{\mu \nu}''(\omega,\beta)$ and the MUC,
\begin{equation} 
\label{ulmsuc1} 
\mathcal{U}_{\mu \nu} = \frac{i}{ \hbar \pi}  \int_{-\infty}^{+\infty} \frac{d\omega}{\omega^2} \tanh^2 \left( \frac{\hbar \omega \beta}{2} \right)\chi_{\mu \nu}''(\omega,\beta) ,
\end{equation}
where the set of perturbations $\{\lambda_\mu\}$ in~(\ref{eq:pert1})
plays the role of the parameters in the derivation of
$\mathcal{U}_{\mu \nu}$. By means of the fluctuation-dissipation
theorem~\cite{Altland2006}, one can further derive an expression
for \Eqref{ulmsuc1} in terms of the dynamical structure factor,
$S_{\mu \nu}(\omega,\beta) = \int_{-\infty}^{+\infty} d t e^{i \omega t}
S_{\mu \nu} (t)$ (i.e. the Fourier transform of the correlation
matrix $S_{\mu \nu}(t) = \braket{\hat{O}_\mu (t) \hat{O}_\nu (0)}$),
which reads
\begin{equation} \label{MUCCorr} \mathcal{U}_{\mu \nu} =
\frac{i}{2 \pi \hbar }  \int_{-\infty}^{+\infty} \frac{d\omega}{\omega^2}
\tanh^2 \left( \frac{\hbar \omega \beta}{2} \right) ( S_{\mu \nu}(\omega,\beta)
- S_{\nu \mu}(-\omega,\beta) ).
\end{equation}
\subsection*{Electrical conductivity and $\bm{n_U}$}
Let's assume now a 2D Fermionic system that presents translational
invariance and let's connect the above formulas to the Uhlmann
number. In the quasi-momentum representation, the
Hamiltonian reads $\HH_{0} = \sum_{\mathbf{k}\in BZ} \funk{\HH}$. If
the system is perturbed by a time-dependent homogeneous electric
field, the Hamiltonian is, up to first order,
\begin{equation} \HH = \HH_0 + \HH_{ext} = \int_{BZ} d^2k
\left( \HH(\mathbf{k}) - \mathbf{J}_{k} \mathbf{A}(t) \right) ,
\end{equation}
where $\mathbf{J}_{k}$ is the electrical current density and
$\mathbf{A}(t)$ is the potential vector. {\color{black} By exploiting standard linear response theory, one is able to link the conductivity, with the derivatives of the $\HH$, as follows 
\begin{align}
\sigma_{\mu \nu}''(\omega,\beta) = \frac{e^2}{\hbar^2 } \sum_{i,j} \int_{BZ}d^2k  \frac{\pi \delta(\omega + \omega_{ij})}{ i \hbar \omega} (p_i - p_j)\times 
 (\partial_{k_\mu}\HH(k))_{ij} (\partial_{k_\nu}\HH(k))_{ji}\qquad \mu, \nu=x,y,
\end{align}
where $\sigma''_{\mu \nu}$ is the dissipative part of the conductivity, defined as $\sigma''_{\mu \nu} (\omega,\beta) := \frac{-i}{2}( \sigma_{\mu \nu} (\omega,\beta) + \sigma_{\nu \mu} ( - \omega,\beta) )$, in terms of the $2\times 2$ conductivity tensor $\sigma_{\mu \nu}$.
By using a procedure similar to that used to
derive~Eq.~(\ref{ulmsuc1}), we are able to calculate the following
formula
\begin{equation} \label{tknnfor} \frac{1}{\pi}\int_{-\infty}^{+\infty}
\frac{d\omega}{\omega}\tanh^2 \left( \frac{\hbar \omega \beta}{2}
\right)  \sigma''_{xy} (\omega,\beta) = - \frac{e^2}{2 \pi \hbar} 
n_U,
\end{equation}
which links the dissipative part of the \emph{dynamical transversal conductivity} $\sigma''_{xy}(\omega,\beta)$ to the Uhlmann
number (\Eqref{nu}).
{ \color{black} 
Exploiting the symmetry properties of the conductivity with respect to $\omega$, and plugging the Kramers-Kroing relations
\begin{equation}
\text{Im} [ \sigma_{\mu \nu} (\omega,\beta) ] =  -\frac{ \omega }{ \pi} \mathcal{P} \int_{-\infty}^{+\infty} \frac{ \text{Re}[ \sigma_{\mu \nu} (\omega')] }{\omega'^2 - \omega^2} d \omega', \qquad \mu, \nu=x,y,
\end{equation}
into~\Eqref{tknnfor}, yields eq.~\eqref{nUvsSigmaR}, which is displayed here for convenience,  
\begin{equation}\label{tknn2}
n_U \frac{q^2}{2 \pi \hbar} = -\int_{-\infty}^{+\infty} d \omega  \tilde{\sigma}_{xy}(\omega,\beta) K_\beta(\omega).
\end{equation}
The above formula shows the dependence of $n_U$ only on $\tilde{\sigma}_{xy}(\omega,\beta)$,
 the real, antisymmetric part of the dynamical transversal conductivity, which can be calculated, following a similar procedure as in \cite{Qi2006}, as 
\begin{align}\label{eq:TCond}
\tilde{\sigma}_{xy}(\omega,\beta):=\frac{\sigma_{xy}^R(\omega,\beta)-\sigma_{yx}^R(\omega,\beta)}{2}&=-\frac{e^2}{\hbar}\frac{1}{(2\pi)^2}\mathcal{P}\int_\textrm{BZ}d^2k \frac{\omega_k^2}{\omega^2_k-\omega^2}\tanh{\left(\frac{\beta \hbar \omega_k}{2} \right) } F^B_{xy},
\end{align}
weighted by the kernel $K_\beta(\omega)$. The latter is defined as }
\begin{equation}
K_\beta(\omega) := \frac{1}{\pi^2} \int_{-\infty}^{+\infty} d \omega' \frac{\tanh^2\left( \frac{\hbar \omega' \beta}{2} \right)}{\omega'^2 - \omega^2}=\left\{\begin{array}{ll}
\frac{1}{i\pi^{3}}\frac{\Psi^{(1)}\left(\frac{1}{2}-\frac{i \hbar  \beta \omega}{2 \pi }\right)-\Psi^{(1)}\left(\frac{1}{2}+\frac{i \hbar  \beta \omega}{2 \pi }\right)}{\omega}& \omega\neq0 \\[8pt]
-\frac{\hbar \beta}{\pi^{4}} \Psi^{(2)}\left(\frac{1}{2}\right)= \frac{14 \hbar \beta}{\pi^{4}} \zeta(3)&\omega=0 \end{array}\right.
\end{equation}
where $\Psi ^{(n)}(z)$ is the n-th poly-gamma function, defined as $\Psi^{(n)}:=\frac{d^{n+1}}{dz^{n+1}} \ln \Gamma[z]$, and $\zeta(z)$ is the Riemann zeta function.
One can demonstrate that $K_\beta(\omega)$ is a probability density function over the frequency domain $\omega\in\mathbb{R}$, i.e. that $K_\beta(\omega)\ge 0$, $\forall \omega,\beta \in \mathbb{R}$ and $\int_{-\infty}^{\infty}d\omega K_\beta(\omega)=1$. In particular,
\begin{align*}
\lim_{\beta\to \infty} K_\beta(\omega) =\delta(\omega),
\end{align*}
showing that eq.~\eqref{tknn2} reduces to the TKNN formula in the zero-temperature limit.\\
Moreover, the probability distribution $K_\beta(\omega)$ is symmetric, peaked at $\omega=0$, and approximately non-vanishing only within a frequency band $\omega\in\{-\Delta\omega,\Delta\omega \}$ of width  $\Delta \omega \simeq \frac{10}{\hbar \beta}$, which provides most of the contributions (about $92\%$) to the integral in~eq.~\eqref{tknn2}.
This shows that $n_U$ can be calculated as a weighted average of the real antisymmetric part of the dynamical transverse conductivity, with a dominant contribution due to the static conductivity, which grows as $1/T$ as temperature decreases.  
}

\section*{Conclusions and outlook}
In this work, we studied two prototypical models of TI and
tested the behaviour of the Uhlmann number against the topological
features of these models at non-zero temperature. {\color{black} We demonstrate the connection of the Ulhmann number to experimentally accessible quantities such as susceptibility and transverse conductivity, and derive a generalised TKKN formula. We investigated the implications of the above formula in both TI models. Our results suggests no indication of temperature-driven topological phase
transitions, nor any actual phase transition at finite-temperature, in both models.}
Instead, we have found that the temperatures smooths out the transition
between regions of zero-temperature topological order. Moreover, we observed
an interesting non-monotonic behaviour of the \emph{Uhlmann number} $n_U$ in the QWZ model,
which can be ascribed to a thermal activation of topological
features for systems which are topologically trivial at zero
temperature. {\color{black} We found that this effect is consistent with the 
appearance of the van Hove singularities in the dynamical conductivity. 
We foresee the possibility of extending the present analysis beyond uncorrelated models~\cite{Yoshida2012,Yoshida2016}.}
%
\bibliography{ref}

\section*{Acknowledgements}
This work was supported by the Government of the Russian Federation
through Agreement No. 074-02-2018-330 (2), and partially by the
Ministry of Education and Research of Italian Government.
\section*{Author contributions statement}
All authors conceived the idea. L.L. and A.C. carried out
calculations, wrote numerical codes and made graphs. All authors
interpreted and explained results. All authors contributed to review
the manuscript.
\section*{Additional information}
%

\textbf{Competing interests:} The authors declare no
competing interests.

\end{document}